# A Geometric Analysis of Time Series Leading to Information Encoding and A New Entropy Measure

Kaushik Majumdar, Systems Science and Informatics Unit, Indian Statistical Institute, 8th Mile, Mysore Road, Bangalore 560059, India; E-mail: kmajumdar@isibang.ac.in

Srinath Jayachandran, Systems Science and Informatics Unit, Indian Statistical Institute, 8th Mile, Mysore Road, Bangalore 560059, India; E-mail: sri9s@yahoo.in

**Abstract**: A time series is uniquely represented by its geometric shape, which also carries information. A time series can be modelled as the trajectory of a particle moving in a force field with one degree of freedom. The force acting on the particle shapes the trajectory of its motion, which is made up of elementary shapes of infinitesimal neighborhoods of points in the trajectory. It has been proved that an infinitesimal neighborhood of a point in a continuous time series can have at least 29 different shapes or configurations. So information can be encoded in it in at least 29 different ways. A 3-point neighborhood (the smallest) in a discrete time series can have precisely 13 different shapes or configurations. In other words, a discrete time series can be expressed as a string of 13 symbols. Across diverse real as well as simulated data sets it has been observed that 6 of them occur more frequently and the remaining 7 occur less frequently. Based on frequency distribution of 13 configurations or 13 different ways of information encoding a novel entropy measure, called semantic entropy (E), has been defined. Following notion of power in Newtonian mechanics of the moving particle whose trajectory is the time series, a notion of information power (P) has been introduced for time series. E/P turned out to be an important indicator of synchronous behaviour of time series as observed in epileptic EEG signals.

1. Introduction

A time series is a collection of observations indexed by the date of each observation (Hamilton 1994, p. 25; Lin et al. 2012, Chapter 28).This is for slowly evolving discrete time series chronicling social events. There are continuous time series like the ECG signal of the heart (Richman and Moorman 2000) or the EEG signals of the brain. In general, any time domain signal is a time series. The term *signal* is generally applied to something that conveys information (Shannon 1948; Oppenheim et al. 1999). Time series are also studied for their underlying information content. In signal processing transformations of the time domain signals to domains other than the time are quite popular. In contrast, time series analyses in economics and other social sciences are more predominantly performed in the time domain itself, in which how the amplitude of the series is varying with respect to time is all that matters. That is why difference operations with respect to time are of such fundamental importance in time series analysis (Hamilton 1994). We too will analyze time series here by difference operations alone.

In this work, probably for the first time ever, we bring Newtonian mechanics to time series analysis. One simple way to do that is to visualize or model a time series as the trajectory of a moving particle. Intuitively, trajectory of the motion of a particle is a displacement function of time and therefore should be a time series. The converse, that is, any time series can be modelled as the trajectory of a moving particle needs a suitable mathematical definition of time series, which we are going to propose here. Introduction of Newtonian mechanics offers novel insights into the geometry of a time series. The geometry of the time series uniquely represents the time series and contains its meaning or the semantic information (Floridi 2015) encoded into it. How geometry is shaped therefore gives the semantic information embedding or encoding into the time series. For example, seismograph signals (time vs. amplitude of rhythms of earth surface) contain information about earth quakes as typical shapes or the shape of the sensex time series tells about the trends of the stock market.

Geometry of time series has been explored in Physics (Packard et al. 1980), in Information Theory (Amari and Nagaoka 2007, see Chapter 5), in Computer Science (Frank et al. 2013), in Finance (Evertsz 1995) and in other areas. Data mining in time series is often performed by sampling points from the time series and constructing new geometric shapes out of them (for an excellent review see Fu 2011). In many empirical analyses of time series geometric features are studied without any concrete theoretical framework. A good example is visual analysis of biomedical signals by clinicians. We have shown here that it is possible to express a discrete time series as a string of finite number of symbols, each representing a distinct elementary geometric local shape in the time series. In other words, it is possible to process a time series as an expression in a natural language by means of natural language processing algorithms, which are already well developed (Manning and Schutze 1999; Berger et al. 1996). One key feature of symbolic representation is transition from one symbol to the next. In general, a time series is a stochastic process and transition from one time point (a random variable) to the next (another random variable) has a transition probability. Similarly, there is a transition probability from one symbol to the next.

Carrying forward the infinitesimal neighborhood shape analysis to continuous time series as well we have identified as many as 29 different shapes or configurations, of which we didn't find any reference in the literature. In fact the shape analysis of a continuous time series has not gotten almost any attention outside the realm of dynamical systems theory (Smale 1967). Here time series are the phase space trajectories that are differentiable one or more times on a compact manifold. Also these time series can revisit the same point multiple times, which we will exclude to keep conformity with economic time series and signals. A somewhat converse problem of constructing the phase space geometry from time series has also been extensively studied (Packard et al. 1980).

One important aspect of a time series is how complex it is. Here complex means how random or unpredictable the time series is. This is measured by entropy – the higher the entropy the more random and less predictable the time series is. Entropy has been measured in time series in many different ways (Acharya et al. 2012). Different entropy measures are suitable for different purposes. In this work we have shown that any discrete time series can be decomposed into precisely 13 different 3-point configurations or geometric shapes. Frequency distribution of those configurations gives a natural measure of entropy, which we call semantic entropy. One pivotal aspect of time series analysis is power spectral estimates, which describe the second-order statistics of the process (Ulrych and Clayton 1976). Here we have introduced the concept of information power based on third-order statistics.

Study of synchronization of multiple time series is of central importance in many fields (Arenas et al. 2009). Epileptic seizures are known to be a hyper-synchronous phenomenon (Fisher et al. 2005). Testing on epileptic seizure data we have observed semantic entropy / information power has the potentiality to indicate synchronization among time series in some systems. We have discussed about the role of the quantity E/(P*K) in synchronization of multiple time series, where E is semantic entropy, P is information power and K is coupling strength. Introduction of K has been inspired by the Kuramoto model of coupled oscillators (Acebron et al. 2005).

## 2. Theoretical Background

*2.1 Model*

There is always an underlying dynamical system (that is, a system which changes with respect to time (Hirsch 1984)) behind generation of a time series (Tong 1990). A dynamical system is represented as $s'(t) = f(s(t))$, where $f$ is a function. As a solution or output of this system we obtain the time series $s(t)$. Clearly, $s(t)$ is at least piecewise first order differentiable. Intuitively this means the graph $(t, s(t))$ can be drawn on a surface by tracing with a mechanical marker along the tangent at point $t$ like the way, say, a seismograph is generated (Fig 1). The amplitude of a one dimensional time series can vary in any manner according to the underlying system generating the time series. So, the ball in Fig 1 that is generating the time series can have only one degree of freedom and that is along the ordinate. It is clear that if the ball moves with a uniform velocity, $s(t)$ will be a monotonic function, whose graph will be a straight line through the origin. In general $s(t)$ will not be a straight line as in Fig 1 and in that case the ball in Fig 1 cannot move with uniform velocity. It must have acceleration.

Notice that the ball in Fig 1 is nothing but a particle moving with one degree of freedom (along the ordinate), whose trajectory is the graph of the time series $s(t)$. Any time series can be generated by controlling the motion of the particle. In general the particle is moving with acceleration and therefore must be in a force field and its motion has only one degree of freedom. So, any one dimensional time series can be modelled as the trajectory of a particle moving in a force field with one degree of freedom. Here we have slightly abused the notation. Actually, trajectory of the particle is the graph of the time series, not the time series itself. However, this should hardly create any confusion.

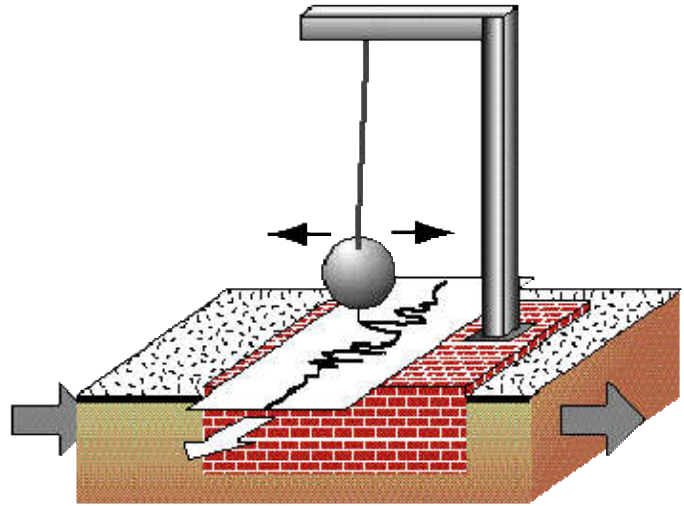

Fig 1. A seismogram is generating a seismograph $(t, s(t))$ as a geological time series. Seismograph is a function of amplitude of earth surface's rhythm (due to the internal dynamics of the planet) over time (http://www.colorado.edu/physics/phys2900/homepages/Marianne.Hogan/graphs.html).

Since the continuous time series $s(t)$ is generated by a particle moving in a force field, the acceleration $s''(t)$ must exist, where $'$ denotes derivative with respect to time. Existence of double derivative at any time $t$ is too stringent a condition. We will therefore try to relax it to the extent that any time series can still be modelled as the trajectory of a moving particle. But first of all we need a precise mathematical definition of time series.

**Definition 1:** A *time series* is a function $s : \Re^+ \cup \{0\} \to \Re$, where $\Re$ is the set of real numbers and $\Re^+$ is the set of positive real numbers, satisfying the following properties.

(1) $s$ is bounded,

(2) $s''$ exists in any compact subset $\square$ of $\Re^+ \cup \{0\}$, except at most at a finite number of points in $\square$.

**Definition 2:** A continuous function $f : \Re \to \Re$ is *traceable* at a point $x \in \Re$ if and only if both left derivative and right derivative of $f$ at $x$ exist.

Clearly, traceable and differentiable are very closely related concepts. If a function is differentiable at a point it must be traceable at that point, but not the other way round. A function to be differentiable at a point left derivative and right derivative of the function at that point must exist. This implies the function must have to be traceable at that point. The function to be differentiable at that point the two derivatives must be equal. Traceability does not have any provision for this condition.

Physically, traceable means the graph of the time series $s(t)$ can be traced from $(t, s(t))$ to $(t+h, s(t+h))$ by a mechanical device all along the graph (that is, without taking off the device from the graph), $h$ is an arbitrarily small real number. When $h$ is negative tracing is from the left and when positive tracing is from the right. This is the way paper ECG (Kao et al. 2001) or paper EEG is traced or recorded. Clearly, the seismograph of Fig 1 has been traced in the same way by the seismogram.

Condition (2) of Definition 1 is a sufficient condition for a time series to be traceable. Any time series is differentiable in a finite interval except at most at a finite number of points. Also $s$ can have at most a finite set of points of discontinuities in the above mentioned interval and all these discontinuities will be jump discontinuities by (1) of Definition 1. Since all time series of real life are of finite duration, the above assertions hold for all real life time series. In order to be the trajectory of a moving particle in a force field a time series must have to be traceable, which is ensured because $s(t)$ is the solution of $s'(t) = f(s(t))$. Also for the force field to exist $s''(t)$ must exist, which is ensured by the condition (2) of Definition 1. It also guarantees that the points where double derivative does not exist are at most finite in number and therefore isolated. In between any two such isolated points $s''(t)$ exists and the particle can trace the time series moving under the force.

**Definition 3:** A point where in a time series the double derivative does not exist will be called *break point*.

**Definition 4:** The point in a time series where the double derivative exists will be called *smooth point*.

A time series is made up of only a finite number of break points and uncountably infinitely many smooth points. Clearly, a time series can be expressed as a union of double differentiable segments, each of which is defined on a maximal open interval of $\Re^+ \cup \{0\}$, where the time series at the boundary point between two adjacent segments is a break point. In each of the maximal open intervals, in which the time series is made up of smooth points only, it can be modelled as the trajectory of a particle moving in a force field with one degree of freedom (Please see the starting paragraph of 2.1). The force is given by $s''(t)$, assuming the mass is unit.

The rate at which work is done by the particle during its motion is

$$P(s(t)) = \frac{d^2 s(t)}{dt^2} \frac{ds(t)}{dt}, \qquad (1)$$

where P stands for power. $P(s(t))$ in (1) gives the power of the particle (as defined in Newtonian mechanics) as the rate at which it works or dissipates its kinetic energy. Now, where does this energy go? This energy is spent to create the trajectory of the particle or the time series. In other words, this energy is spent to create the time series, in which the kinetic energy of the particle is transformed into information to be embedded in it in terms of its shape. The shape of the time series gives its meaning and therefore we can propose the following

**Definition 5:** The meaning or the interpretation of a time series is called its *semantic information*. In this work by semantic information we will mean the interpretation of a time series in terms of its shape in the time domain.

It is pertinent to note here that semantic information is a well studied area (Carnap and Bar-Hillel 1952), but there is no universally accepted definition for it (Floridi 2005). We therefore go by the so called standard interpretation of semantic information (Floridi 2005) for time series, which has been encapsulated in Definition 5. Also Definition 5 is commensurate with the *general definition of information* (GDI) in (Floridi 2015). Putting equation (1) and Definition 5 together we conclude that $|P(s(t))|$ is the rate at which semantic information is being encoded in a time series $s(t)$ at time $t$.

**Definition 6:** The total *information power* $P_I(s(t))$ of a time series $s(t)$ in an interval $(a,b)$ is given by $P_I(s(t)) = \int_a^b |P(s(t))| dt$.

The information power is not affected by the presence of break points in $(a,b)$. We will see how this information power of a time series is related to synchronization of the component time series of that time series (when the time series $s(t)$ is made up of several other time series). Information power will be discussed again in subsection 3.2.

*2.2 Discretization*

Since a time series will have to be computed anyway by electronic computers, it has to be discretized (in old days analog signals used to be processed by passing the signal through special purpose electronic circuits, but now one single computer can do huge varieties of digital signal processing only by executing particular algorithms for particular purposes). Once the time series is discretized, derivative is to be replaced by difference operation. Here we will assume that for discretization the continuous-time time series has been sampled at equal intervals. However for unequal interval sampling all the arguments will hold only with slight modification for the unequal time intervals. Throughout this work we will denote continuous-time time series by $s(t)$ and discrete-time time series by $s[n]$. $P(s(t))$ will be replaced by $P(s[n])$, where we define the *first order difference* $s'[n]$ either as $s[n]-s[n-1]$ (*backward difference*) or as $s[n+1]-s[n]$ (*forward difference*).

Note that $P(s[n])$ is a product of first difference and second difference operations on $s[n]$ in the neighbourhood of $n$, and $P(s[n])$ needs at least a 3-point neighbourhood to perform the double difference operation, because after the double difference operation only 1 point out of the 3-point neighbourhood will survive. The outcome of the P-operator in that neighbourhood will have to be attributed to that surviving point. In Fig 2 the isolated neighbourhood consisting of time points $n-1, n$, and $n+1$ has been considered. $s''[n]$ should be attributed to $n$ and therefore $n$ must survive after the double difference operation. Note that if we follow the backward difference convention for both the first difference and the second difference operations the value of the double difference operation will have to be attributed to $n-1$ and for following the forward difference convention for the first and the second difference operations the value will have to be attributed to $n+1$. In order to attribute the double difference to $n$ we must follow one convention for the first difference and the other convention for the second difference. In this work we will follow backward difference for the first difference operation and forward difference for the second difference operation.

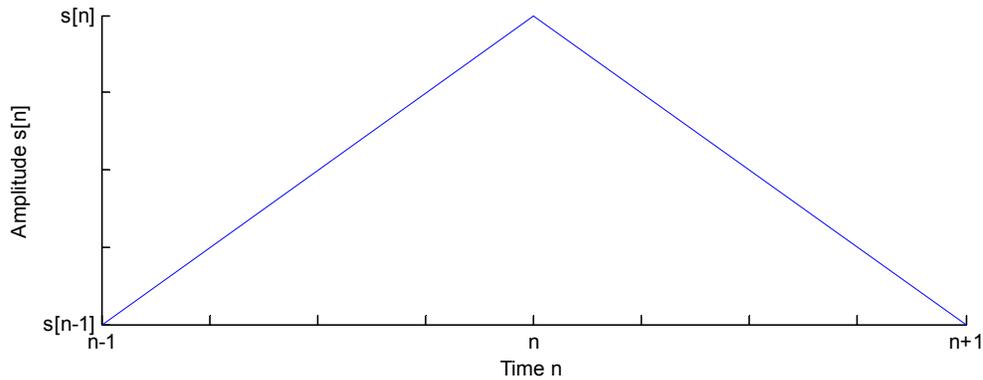

Fig 2. Time (n) vs. amplitude (s[n]) plot of the smallest neighborhood of s[n], keeping s[n] as an interior point. Left product of P(s[n]) is P(s[n-]) = s''[n]*s'[n] and right product of P(s[n]) is P(s[n+]) = s''[n]*s'[n+1], where s'[n] = s[n] – s[n-1] and s''[n] = s'[n+1] – s'[n].

In Fig 2 it has been elaborated how P-operator value on $s[n]$ at $n$ can be attributed as a product of second difference and backward difference or second difference and forward difference at the same point.

**Definition 7:** We define *left product* of P-operator on $s[n]$ at $n$ as $P(s[n-]) = s''[n] * s'[n]$, where $s'[n] = s[n] - s[n-1]$ and $s''[n] = s'[n+1] - s'[n]$.

**Definition 8:** We define *right product* of P-operator on $s[n]$ at $n$ as $P(s[n+]) = s''[n] * s'[n+1]$, where $s'[n+1] = s[n+1] - s[n]$ and $s''[n] = s'[n+1] - s'[n]$.

How P-operator on $s[n]$ at $n$ is changing (or not changing) sign from left product to right product determines the shape of the 3-point neighborhood consisting of $n-1$, $n$ and $n+1$ as we will see very soon. It is also interesting to note that given the sign hierarchy as defined in Definition 9, P-operator can change sign from left product to right product only according to the sign hierarchy (please see Theorem 1). That is, P-operator can never change sign from positive to zero, zero to negative or from positive to negative.

**Definition 9:** By *sign hierarchy* we mean $- < 0 < +$.

**Lemma 1:** P-operator cannot change sign from $0$ to $-$.

**Proof:** Let the contrary be true. Then for some $n$ $P(s[n-]) = 0$ and $P(s[n+]) < 0$.
$P(s[n-]) = 0 \Rightarrow s''[n] * s'[n] = 0$, that is, $s'[n] = 0$, because $s''[n] \neq 0$ as $s''[n] * s'[n+1] < 0$.
$s'[n] = 0 \Rightarrow s[n] = s[n-1]$ ........ (a).
$s''[n] * s'[n+1] < 0 \Rightarrow s'[n+1] > 0$ and $s''[n] < 0$, or, $s'[n+1] < 0$ and $s''[n] > 0$. If the first is true then $s[n+1] < s[n]$ ....... (b) and $s[n+1] + s[n-1] > 2 * s[n]$ ...... (c). (a) and (b) together imply $s[n+1] + s[n-1] < 2 * s[n]$, which is contradictory to (c). Therefore, $s'[n+1] > 0$ and $s''[n] < 0$ cannot hold. Similarly, it can be shown that $s'[n+1] < 0$ and $s''[n] > 0$ also cannot hold. Therefore, $P(s[n-]) = 0$ and $P(s[n+]) < 0$ cannot hold. This completes the proof.

**Lemma 2:** P-operator cannot change sign from $+$ to $0$.

**Proof:** Similar.

**Lemma 3:** P-operator cannot change sign from $+$ to $-$.

**Proof:** Similar.

**Theorem 1:** P-operator can change sign from left product to right product only according to the sign hierarchy described in Definition 9.

**Proof:** Lemmas 1, 2 and 3 demonstrate that it is not possible for the left product of P-operator to have higher sign value as enlisted in Definition 9 and the right product to have a lower sign value. So, the left product of P-operator can only have lower sign value and the right product can only have a higher sign value. This completes the proof.

**Lemma 4:** The left product of P-operator is zero and the right product of P-operator is also zero can happen only in the following three ways:

1. $s'[n] > 0$, $s''[n] = 0$, $s'[n+1] > 0$.
2. $s'[n] < 0$, $s''[n] = 0$, $s'[n+1] < 0$.
3. $s'[n] = 0$, $s''[n] = 0$, $s'[n+1] = 0$.

**Proof:** In the first place it appears that both left and right products of P-operator can be zero in 11 different ways. This is because, if $s''[n] = 0$, then left product can take all three different signs and also the right product can take all three different signs independently of each other, and therefore account for 3*3 = 9 cases. When $s'[n] = 0$ and $s'[n+1] = 0$, $s''[n]$ can either be negative or positive. So, there are a total of 9 + 2 = 11 number of different ways this can be vanishing of both left and right products can be achieved. However, a deeper analysis will reveal except the above three cases all other cases are ruled out. For example, let us examine the case $s'[n] > 0$, $s''[n] = 0$, $s'[n+1] < 0$.

Since $s'[n] > 0$, $s[n] > s[n-1]$ ...... (a). $s''[n] = 0 \Rightarrow s[n-1] + s[n+1] = 2 * s[n]$ ...... (b).

$s'[n+1] < 0 \Rightarrow s[n+1] < s[n]$ ...... (c). (a) and (c) together give $2 * s[n] > s[n-1] + s[n+1]$,

which is contrary to (b). So $s'[n] > 0$, $s''[n] = 0$, $s'[n+1] < 0$ cannot hold. With very similar arguments all the 8 possibilities can be ruled out. This completes the proof.

**Theorem 2:** The 3-point neighbourhood of $s[n]$ consisting of $n-1$, $n$ and $n+1$ can have precisely 13 distinct geometric configurations enlisted in Table 1.

Table 1

Configurations the neighbourhood of $s[n]$ consisting of points $n-1$, $n$ and $n+1$ can take alongside the corresponding sign change from left product to right product of P-operator.

| Sign change | s'[n] | s''[n] | s'[n+1] | 3-point configuration | Configuration number | Empirical observation |
|---|---|---|---|---|---|---|
| -- | - | + | - | | 1 | Statistically more abundant |
|  | + | - | + | | 2 | |
| ++ | + | + | + | | 3 | |
|  | - | - | - | | 4 | |
| -+ | - | + | + | | 5 | |
|  | + | - | - | | 6 | |
| 00 | + | 0 | + | | 7 | Statistically less abundant |
|  | - | 0 | - | | 8 | |
|  | 0 | 0 | 0 | | 9 | |
| 0+ | 0 | + | + | | 10 | |
|  | 0 | - | - | | 11 | |
| -0 | + | - | 0 | | 12 | |
|  | - | + | 0 | | 13 | |

**Proof:** Left product of P-operator involves two factors. Right product of P-operator also involves two factors. In both cases $s''[n]$ is common. So there are only three factors involved in both the products taken together. Each can take sign in three different ways. So there are $3^3 = 27$ different possibilities.

Lemma 1 rules out two possibilities. Since the right product is not zero $s''[n]$ cannot be zero. So for sign change from zero to negative $s'[n]$ must be zero. Either $s''[n] > 0$ and $s'[n+1] < 0$ or $s''[n] < 0$ and $s'[n+1] > 0$. Both of these are ruled out by Lemma 1.

Similarly, Lemma 2 and Lemma 3 rule out two possibilities each. Six possibilities are ruled out this way. Lemma 4 rules out eight more. Altogether fourteen possibilities are ruled out and therefore 27 – 14 = 13 possibilities survive, all of which have been listed in Table 1. This completes the proof.

**Lemma 5:** When P-operator changes sign from negative to positive either a peak (configuration number 6 in Table 1) or a trough (configuration number 5 in Table 1) appears in $s[n]$.

**Proof:** A peak is simply a local maximum and a trough is a local minimum. If $s[n]$ has a peak at $n$, $s[n]$ must increase from $n-1$ to $n$ and must decrease from $n$ to $n+1$. So, $s'[n] > 0$ and $s'[n+1] < 0$. It is easy to verify that $s''[n] < 0$. The left product of $P(s[n])$ is negative and the right product is positive. Similarly it can be shown that for a trough of $s[n]$ at $n$ $s'[n] < 0$, $s'[n+1] > 0$ and $s''[n] > 0$, leading to left product negative and right product positive. This completes the proof.

Lemma 5 proves that to occur a peak or a trough in a time series sign change must take the longest route, that is, from negative to positive. It is clear from Table 1 that for all other configurations to occur sign change takes place from negative to zero or from zero to positive or no sign change at all. In this sense a peak or a trough is the most significant event in a time series. For the rest of the paper we will refer to a configuration by its configuration number as listed in Table 1.

We have calculated the frequency distribution of the 13 configurations across diverse real and simulated data sets. This includes astronomical data, speech data from different speakers, depth and scalp EEG signals from human subjects recorded in different hospitals and laboratories in Europe, USA and India, rat EEG data, simulated noise signals by different methods (white Gaussian noise in MATLAB, shifted surrogate of real signals, shifted surrogate with power spectrum preserved), real and simulated extracellular potential, simulated EEG from neural mass model, etc. The sampling frequency ranged from 200 Hz to 800 kHz. We observed peaks (configuration 6) and troughs (configuration 5) occurred in all data sets with the highest frequency, configurations 1 through 6 occur quite abundantly across diverse data sets and configurations 7 through 13 occur very sparsely across data sets. There are sufficiently large data sets in which 7 through 13 do not occur at all.

*2.3 Transition*

It is clear that a discretized time series can be expressed as a string of 13 configurations or symbols. However, not every symbol follows or precedes all symbols. For any two successive symbols in a string the first symbol's last two points must be the same as the second symbol's first two points. Among the 13 configurations 59 different transitions are possible (Fig 3). In Fig 3 a 13 x 13 transition matrix has been presented. The $ij$ th entry in the matrix is associated with the transition from configuration $i$ (in Table 1) to configuration $j$, where $i, j \in \{1,....,13\}$. If the $ij$ th entry is zero then no transition is possible from configuration $i$ to configuration $j$. Otherwise the $ij$ th entry is a number between 0 and 1. This number indicates the frequency of occurrence of the $ij$ th transition in a time series window. Therefore summation of all the 59 valid transition entries will be 1.

|    | 1     | 2      | 3     | 4      | 5      | 6     | 7       | 8       | 9      | 10    | 11    | 12     | 13     |
|----|-------|--------|-------|--------|--------|-------|---------|---------|--------|-------|-------|--------|--------|
| 1  | 0.046 | 0      | 0     | 0.044  | 0.043  | 0     | 0       | 0.0007  | 0      | 0     | 0     | 0      | 0.019  |
| 2  | 0     | 0.047  | 0.044 | 0      | 0      | 0.043 | 0.0007  | 0       | 0      | 0     | 0     | 0.019  | 0      |
| 3  | 0     | 0.076  | 0.044 | 0      | 0      | 0.018 | 0.0007  | 0       | 0      | 0     | 0     | 0.011  | 0      |
| 4  | 0.078 | 0      | 0     | 0.046  | 0.018  | 0     | 0       | 0.0008  | 0      | 0     | 0     | 0      | 0.011  |
| 5  | 0     | 0.018  | 0.043 | 0      | 0      | 0.015 | 0.0004  | 0       | 0      | 0     | 0     | 0.009  | 0      |
| 6  | 0.018 | 0      | 0     | 0.044  | 0.015  | 0     | 0       | 0.0004  | 0      | 0     | 0     | 0      | 0.009  |
| 7  | 0     | 0.0007 | 0.0007| 0      | 0      | 0.0005| 0.00001 | 0       | 0      | 0     | 0     | 0.0002 | 0      |
| 8  | 0.0007| 0      | 0     | 0.0007 | 0.0005 | 0     | 0       | 0.00001 | 0      | 0     | 0     | 0      | 0.0002 |
| 9  | 0     | 0      | 0     | 0      | 0      | 0     | 0       | 0       | 0.0019 | 0.006 | 0.006 | 0      | 0      |
| 10 | 0     | 0.011  | 0.018 | 0      | 0      | 0.0094| 0.0002  | 0       | 0      | 0     | 0     | 0.0060 | 0      |
| 11 | 0.012 | 0      | 0     | 0.019  | 0.0089 | 0     | 0       | 0.00025 | 0      | 0     | 0     | 0      | 0.0059 |
| 12 | 0     | 0      | 0     | 0      | 0      | 0     | 0       | 0       | 0.0060 | 0.016 | 0.023 | 0      | 0      |
| 13 | 0     | 0      | 0     | 0      | 0      | 0     | 0       | 0       | 0.0061 | 0.023 | 0.017 | 0      | 0      |

Fig 3. An example of the 13 x 13 stochastic transition matrix. Transition is from row to column. 0 indicates forbidden transition. A nonzero entry signifies the frequency distribution of the transition over a time series window. Symbol numbers are as in Table 1.

Depending on the window size (to be decided by trial and error) this stochastic transition matrix can have important information about the pattern of the discrete time series. In Table 1 we have mentioned about more abundant configurations and less abundant configurations in a time series. This has been checked on wide ranging data sets. There are 6 more abundant configurations and 7 less abundant configurations. Accordingly, the 13 x 13 matrix can be subdivided into 6 x 6, 6 x 7, 7 x 7 and 7 x 6 matrices to study how the transitions take place among and within the more abundant and less abundant configuration classes.

It is clear, by expressing a discrete time series as a string of 13 symbols we are mimicking expressions in natural languages (such as, words and sentences as strings of characters), in which a class of transitions may be more meaningful than the others depending on the system that is generating the time series. Statistical natural language processing is a well developed area (Manning and Schutze 1999). The techniques of natural language processing can be brought to the domain of time series analysis.

### 2.4 Continuous Time Series

We started this section modelling continuous-time time series or simply continuous time series as the trajectory of a moving particle in a force field. Then we went on to analyzing its discretized version. We have again come back to continuous time series to investigate if we can derive here a result like Theorem 2. Fortunately this is feasible by virtue of Definition 1 and Theorem 2.

**Theorem 3:** A continuous time series is made up of at least 29 different configurations.

**Proof:** A continuous time series is made up of break points (Definition 3) and smooth points (Definition 4). At a break point the time series can be either (1) discontinuous with a jump discontinuity, or (2) continuous, but not differentiable, or (3) differentiable, but not double differentiable. By Definition 1 there can be only finite number of break points in any finite interval. Therefore, for jump discontinuities the segment between two successive jump discontinuities can be considered. Union of all such segments will give the time series. From this point onward in this proof we will only consider the continuous segment of the time series.

In the (infinitesimal) neighborhood of a smooth point the double derivative of the time series is either positive, negative or zero. If it is positive, the time series is convex. If the double derivative is negative the time series is concave. If the double derivative is zero the time series is a straight line.

In any finite interval there can be only at most a finite number of break points in a continuous time series. In order to prove it let us take an arbitrary interval $(a,b) \subset \Re^+ \cup \{0\}$. $\overline{(a,b)} = [a,b]$, which is compact by Heine-Borel theorem. By Definition 1 $[a,b]$ and therefore $(a,b)$ must have only at most a finite number of break points. This implies break points are isolated points.

Since break points are isolated points, in an infinitesimal neighborhood of a break point the time series must be double differentiable at all points except at the break point itself. Since the time series is differentiable on both left and right sides of the break point, tangent to the time series on both left and right side of the break point exists. These two tangents can meet at the break point in 10 out of 13 configurations listed in Table 1. Only configurations 7, 8 and 9 are not possible, because these configurations are actually smooth point configurations (double derivative exists at the middle point). In addition there can be a break point where the tangent is perpendicular to the abscissa (Fig 4). 4 distinct configurations are possible (Fig 4).

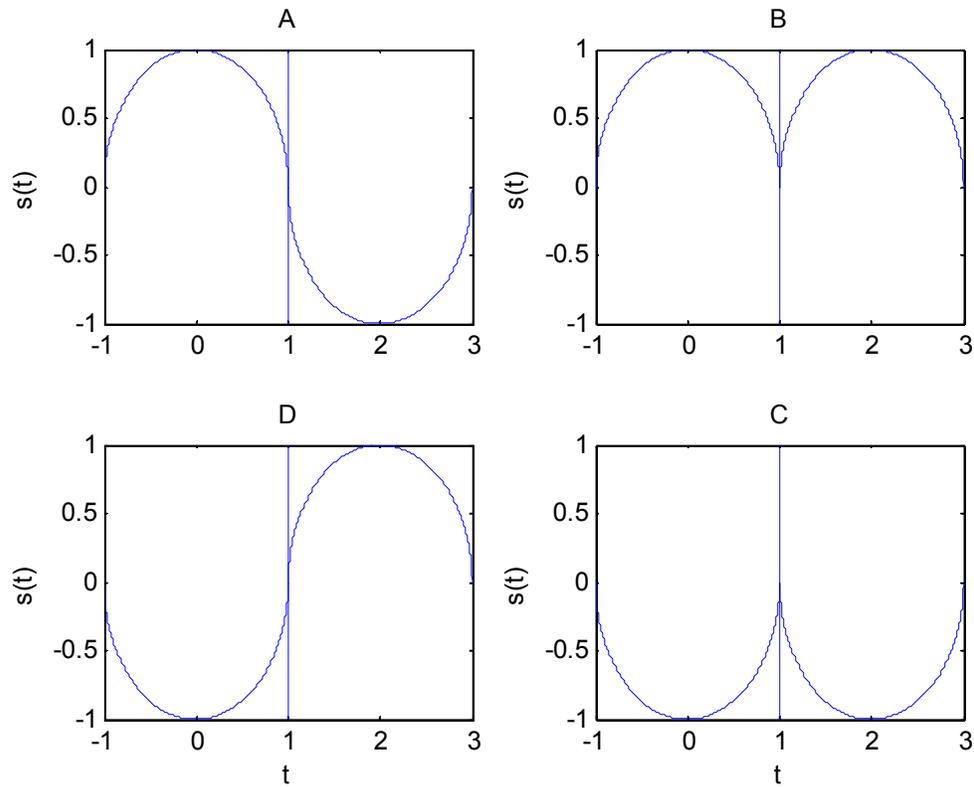

Fig 4. All four configurations of $s(t)$ when there is a vertical tangent to $s(t)$ at $t=1$.

We have already seen that at a smooth point the time series can be convex or concave or a straight line. It can be convex with first derivative negative or positive. It can be concave with first derivative negative or positive. It can be straight line with first derivative negative or positive. Whenever the first derivative is positive the time series is an increasing function of time. Whenever the first derivative is negative the time series is a decreasing function in time. At a smooth point the time series can have one of the 7 configurations listed in Fig 5.

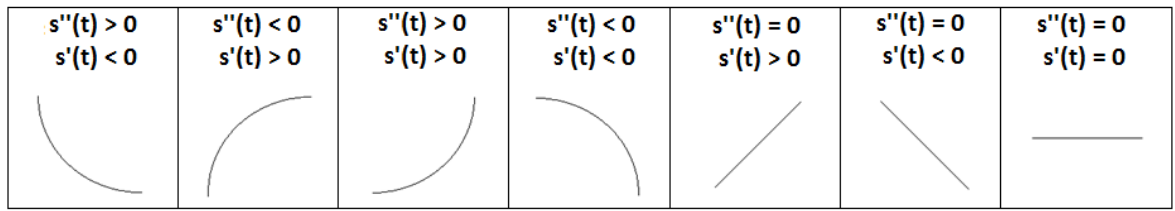

Fig 5. Possible configurations of an infinitesimal neighborhood of a smooth point in any continuous time series with the sign of first and second derivative. Last three configurations are configurations 7, 8 and 9 of Table 1 respectively.

We will call the configurations in Fig 5 as *fundamental configurations*, because any smaller open neighborhood within the configuration will have the same configuration. A break point can occur only at the boundary between two such configurations (in general, boundary between two such configurations can be smooth as well). At a break point it may happen even the first derivative does not exist or the first derivative exists but the second does not exist. In the first case we have 10 configurations in Table 1 plus the 4 configurations of Fig 5 as discussed. In the second case ($s'$ exists but $s''$ does not) the following situation will arise.

Since in a finite interval $s''(t)$ may not exist only at most at a finite number of points, all those points are isolated points. In some left and right neighborhood of such a point $s''(t)$ will exist and therefore $s'(t)$ will be continuous. So, an infinitesimal open neighborhood of $s'(t)$ at $t$ having the break point as an interior point will have one of the 10 configurations in Table 1 plus the first 4 in Fig 5 on either side of the break point as discussed and displayed above. This means the line joining $(t-h_1, s'(t-h_1))$ and $(t, s'(t))$ for $h_1 \to 0$ and the line joining $(t, s'(t))$ and $(t+h_2, s'(t+h_2))$ for $h_2 \to 0$, where both $h_1$ and $h_2$ are positive, will be on the graph of $s'$. In other words, the graph of $s'$ in the infinitesimal neighborhood of $s'(t)$ is represented by the two straight lines for $h_1 \to 0$ (Fig 6) and $h_2 \to 0$.

We will now try to recover $s(t)$ from $s'(t)$. If we integrate $\int_{t-h_1}^{t} s'(x)dx$ we will get the area of the shaded region in Fig 6, which is $m_1 h_1^2 / 2$, where $m_1$ is the slope of the left tangent to $s'$ at $t$ and $h_1 \to 0$. Similarly, on the right side of $(t, s'(t))$ it will be $m_2 h_2^2 / 2$, where $m_2$ is the slope of the right tangent to $s'$ at $t$ and $h_2 \to 0$, $m_2 \neq m_1$. Notice that $h_1$ and $h_2$ are infinitesimal variables. $s(t)$ is of the form $m_1 h_1^2 / 2 + s(t-h_1)$ to the left of $t$ and $s(t+h_2) - m_2 h_2^2 / 2$ to the right of $t$, $m_1 \neq m_2$.

Since the time series $s$ is continuous at $t$ the following equation must hold

$$\lim_{h_1 \to 0} \frac{m_1 h_1^2}{2} + s(t-h_1) = s(t) = s(t+h_2) - \lim_{h_2 \to 0} \frac{m_2 h_2^2}{2}. \tag{2}$$

Since $h_1^2 > 0$ and $h_2^2 > 0$, $s(t) - s(t-h_1)$ and $m_1$ must have same sign, and similarly, $s(t+h_2) - s(t)$ and $m_2$ must have same sign. It is clear from Fig 6 that $m_1 = s''(t-)$ and similarly, $m_2 = s''(t+)$. This means when $s$ is increasing $s'$ is also increasing and when $s$ is decreasing $s'$ is also decreasing. So, on the left of $t$ both $s$ and $s'$ are increasing or decreasing and for that on the right of $t$ both of them are increasing or decreasing. In total there are four possibilities.

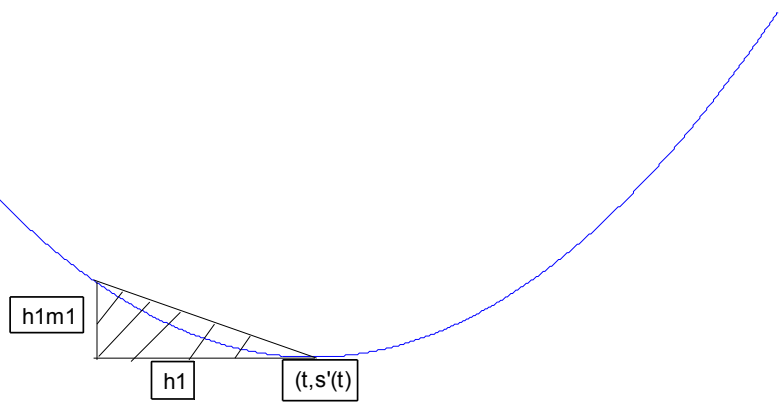

Fig 6. The curve is the graph of $s'$. The arc between $(t-h_1, s'(t-h_1))$ and $(t, s'(t))$ will superpose on the secant for $h_1 \to 0$. $m_1$ is the gradient of the tangent to $s'$ at $(t, s'(t))$. The shaded area will be the integral of $s'(t)$ from $t-h_1$ to $t$ and will give $s(t) - s(t-h_1)$.

The case is still left when $s'$ is continuous at a point, but $s''$ is infinite, that is, when $s'$ will have one of the four configurations of Fig 4. If $s'$ has configurations of Fig 4, $s$ will have configurations of Fig 7. To be precise, plots in Fig 4 have been generated by taking the upper and lower hemispheres of unit circles centered at $(0,0)$ and $(2,0)$. For example, Fig 4(A) has been generated by $s(t) = \sqrt{1-t^2}$ for $-1 \leq t \leq 1$ and $s(t) = -\sqrt{1-(t-2)^2}$ for $1 \leq t \leq 3$. Clearly, $s'(1) = \infty$. Now, if we replace $s(t)$ by $s'(t)$, we get $s'(t) = \sqrt{1-t^2}$ for $-1 \leq t \leq 1$ and $s'(t) = -\sqrt{1-(t-2)^2}$ for $1 \leq t \leq 3$. Solving $s(\theta) = \int_0^\theta \sqrt{1-t^2}\, dt$ for $0 \leq \theta \leq 1$ and $s(\theta) = -\int_1^\theta \sqrt{1-(t-2)^2}\, dt$ for $1 \leq \theta \leq 2$ we get, $s(\theta) = \frac{1}{4}\sin(2\theta) + \frac{1}{2}\theta$ for $0 \leq \theta \leq \frac{\pi}{2}$ and $s(\theta) = -\frac{1}{4}\sin(2\theta) - \frac{1}{2}\theta$ for $-\frac{\pi}{2} \leq \theta \leq 0$, whose simulation gives Fig 7(A). Likewise the other plots of Fig 7 have been generated.

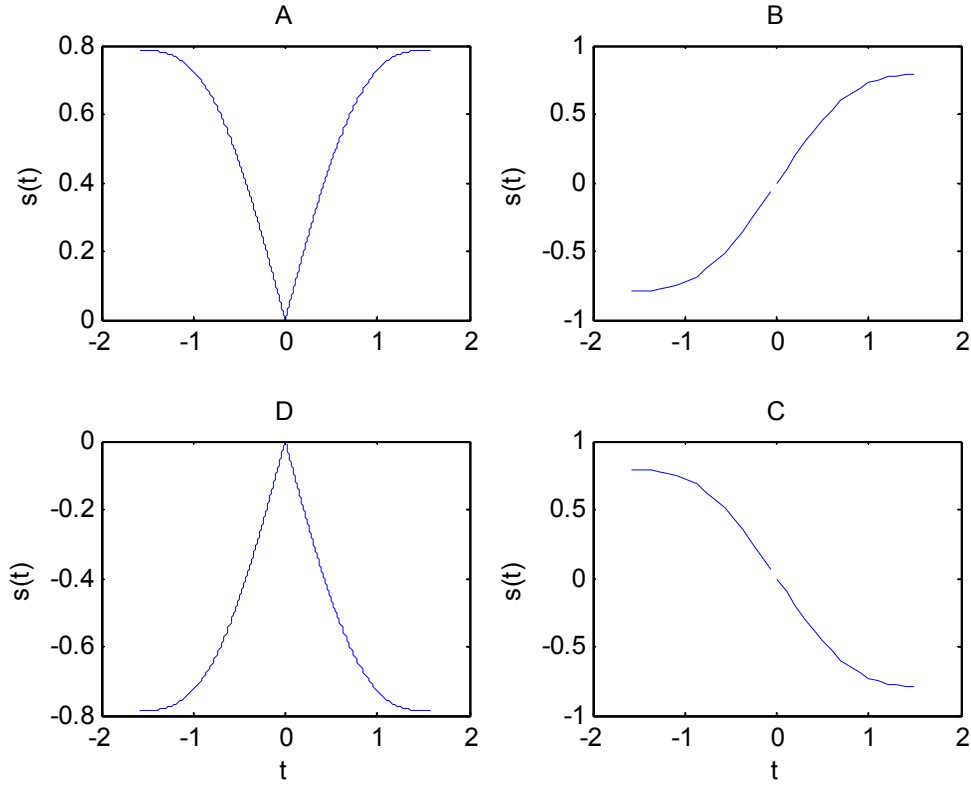

Fig 7. Numerical simulation of $s(t)$ in (A), (B), (C) and (D) provided $s'(t)$ takes the form of Fig 4(A), Fig 4(B), Fig 4(C) and Fig 3(D) respectively for $t \in [0,1] \cup [1,2]$. In all the plots portion of $s'(t)$ for $t \in [1,2]$ in Fig 4 has been mapped on the portion of $s(t)$ in this figure for $t \in [-2,0]$.

By now we have identified 10 (in Table 1) + 4 (in Fig 4) + 7 (in Fig 5) + 4 (in equation (2) or in Fig 6) + 4 (in Fig 7) = 29 different configurations in a continuous time series. So, there are at least 29 configurations or shapes of an infinitesimal neighborhood of a point in a continuous time series. This completes the proof.

**Corollary 1:** A time series $s(t)$ satisfying Definition 1 can have at least 32 different configurations for an infinitesimal neighborhood of a point $t$.

**Proof:** By Definition 1 $s(t)$ is bounded and therefore admits only jump discontinuities. If $s(t)$ is continuous at $t$ then the following must hold for $h > 0$.

$$\lim_{h \to 0} s(t-h) = s(t) = \lim_{h \to 0} s(t+h). \qquad (3)$$

For $s(t)$ to have a jump discontinuity at $t$ at least one of the two equalities in equation (3) must be an inequality. This can happen in three different ways. So, there can be three different jump discontinuities in $s(t)$. By Theorem 3 an infinitesimal neighborhood of a point $t_{s(t)}$ can have at least 29 + 3 = 32 configurations. This completes the proof.

## 3. Application

*3.1 Entropy*

Measure of entropy in a time series gives a mean to estimate the average information content of the time series and therefore regarded as an important measure of a time series (Richman and Moorman 2000, Bandt and Pompe 2002). The 13 configuration decomposition of a discrete time series (Theorem 2) readily gives a measure of entropy in terms of distribution of those configurations in the time series. Since each configuration is the semantic information (Definition 5) about the smallest neighborhood of a point containing that point as an interior point, the entropy based on their distribution should be called semantic entropy.

**Definition 10:** *Semantic entropy* (SE) of a signal of length $N$ is given by $SE = -\sum_{i=1}^{13} p_i \log_2 p_i$, where $p_i = \frac{n_i}{N-2}$ is the frequency distribution of the $i$ th configuration (according to Table 1). $n_i$ is the number of times the $i$ th configuration has occurred in the $N$ point long signal segment.

Note that since P-operator involves double difference, after operating it on the $N$ point long time series an output of a $N-2$ points long time series is produced. We have compared semantic entropy with permutation entropy (Band and Pompe 2002) on focal intracranial EEG signals of epilepsy patients in Freiburg Seizure Prediction Project data base, which is publicly available. Here the permutation entropy goes up and the semantic entropy goes down (Fig 8). So semantic entropy gives a sharper trend. Algorithmically permutation entropy comes closer to semantic entropy than other entropies. Permutation with 3 successive samples of a time series will give 3! = 6 configurations, which are subsumed in the 13 configurations in Table 1. Permutation entropy has been applied on epileptic brain signals during seizure (for a review see Zanin et al. 2012). Entropy by different measures of the EEG signal during seizure goes down (Table 1 of Acharya 2012). Both permutation entropy and semantic entropy go down during the seizure in majority of the cases, but permutation entropy goes up more number of times than semantic entropy goes up. Fig 8 presents an instance where permutation entropy goes up during the seizure, but semantic entropy goes down. Our comparative study between semantic and permutation entropy on 261 focal intracranial EEG channels implanted in 21 patients during recording of 87 seizures has been summarized in the histogram plot of Fig 9.

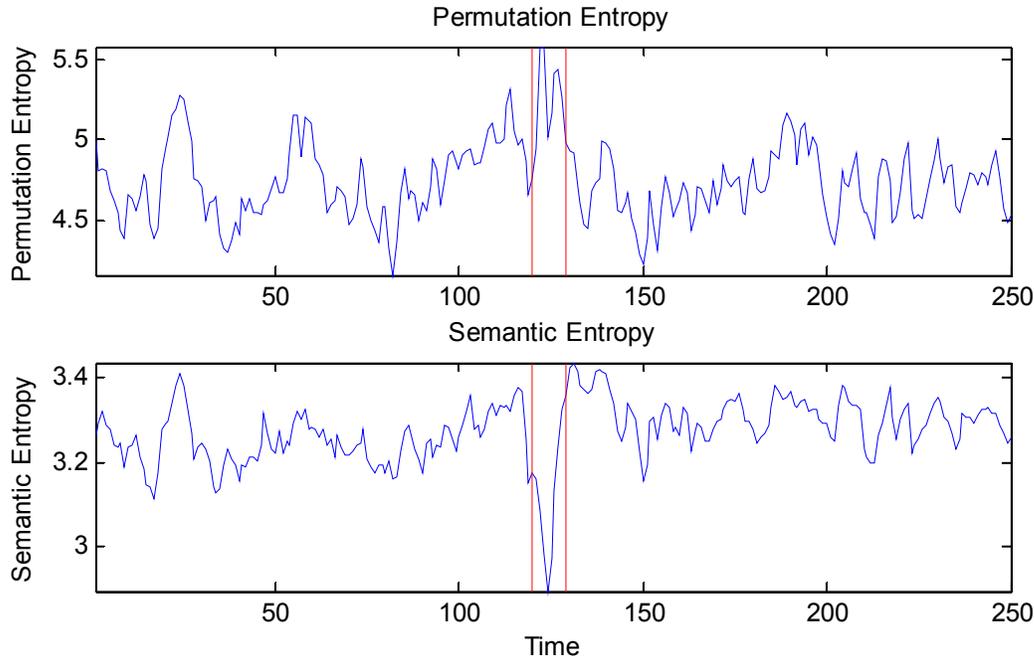

Fig 8. Time versus permutation entropy (top) and semantic entropy (bottom) plot of a human intracranial EEG signal from seizure focus. Vertical lines indicate start and end of an epileptic seizure.

It is clear from Fig 9 that in 46% of all the cases, that is study of signals from 3 focal channels for a duration of one hour which contains one seizure somewhere within that duration, permutation entropy and semantic entropy behaved the same way (in 64% out of the 46% of the cases entropy goes down and in remaining 36% of the cases it does not go down). In 36% of the cases semantic entropy during the seizure goes down, but permutation entropy goes up (like in Fig 8). Usually, pathological cases are signified by reduced entropy (Costa et al. 2002). In the remaining 18% of the cases permutation entropy during the seizure goes down, but semantic entropy goes up. Entropy by a particular measure can go up during a particular seizure, which otherwise goes down in a majority of the cases. However, a measure that gives a more uniform trend across the seizures will be better suited for developing useful algorithms, like automated seizure detection and seizure prediction. Both these areas are very active areas of research with potential healthcare implications for about 1% of the world population estimated to be afflicted with epileptic seizures (Zanin et al. 2012).

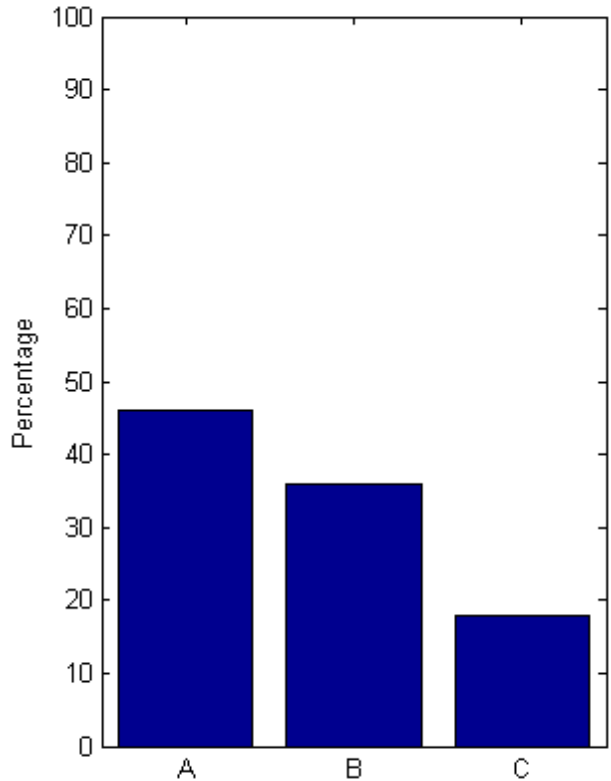

Fig 9. Summary of comparative study between permutation entropy and semantic entropy on 261 focal channels of 21 patients during recording of 87 seizures in Freiburg data set. Histogram A stands for 46% of the cases where both the measures have given same trend. Histogram B stands for 36% of the cases where semantic entropy during the seizure goes down, but permutation entropy goes up. Histogram C signifies the remaining 18% of the cases where during the seizure permutation entropy goes down but semantic entropy goes up during the seizure.

*3.2 Information Power*

Information power has been defined in Definition 6. It has been inspired by the notion of power in Newtonian mechanics and has nothing to do with the spectral power of a time series. Consider time series $s(t) = c$, where $c$ is constant. This will be a flat line parallel to the time axis and carries no information. Information is embedded into the time series $s(t)$ when it fluctuates up and down (along the ordinate) with respect to a fixed position like $s(t) = c$ (fixed ordinate). The more random this fluctuation is the more is the information content of the time series according to Shannon's theory, albeit the meaning of this information or in other words, its semantic content is irrelevant to Shannon's theory. In this work we precisely intend to study this very semantic content of the time series in the form of its geometric shape. Of course different contexts may give different meaning to the same shape of the time series.

How this shape is created can be modelled conveniently by Newtonian mechanics. Consider $s(t) = c$, which is the case when a particle is stationary with respect to the observer. The particle is not doing any 'work' (in the sense of Newtonian mechanics) as it is not making any movements under the influence of a force. The next simplest form the time series can take is $s(t) = ct$ irrespective of whether it is practically meaningful or not. $s(t) = ct$ signifies a displacement $s(t)$ in time $t$ made by a particle moving with a uniform velocity $c$. It will give an unbounded straight line passing through the origin and therefore $s(t) = ct$ will not qualify as a time series according to condition 1 of Definition 1. The only way for $s(t)$ to be an 'informative' time series is to go up and then come down, or to go down and then come up (both the activities within a specified range, because $s(t)$ is bounded) over the time. This is analogous to the case when a particle is displaced from stationarity or from uniform motion by altering its momentum according to the second Newtonian law of motion. Assume the mass of the particle is one. Then the rate of change of momentum of the particle is $\frac{d}{dt}\left(1.\frac{ds(t)}{dt}\right) = \frac{d^2s}{dt^2}$. So, each time the time series is displaced with respect to a stationary position $(s(t) = c)$ it can be thought of being displaced by a 'force' $\frac{d^2s}{dt^2}$ working along the ordinate (therefore, with only one degree of freedom). The particle makes an infinitesimal displacement $ds$ under the influence of the force $\frac{d^2s}{dt^2}$ and thereby executes $\frac{d^2s}{dt^2} * ds$ amount of work. Or, in other words, the particle dissipates $\frac{d^2s}{dt^2} * ds$ amount of kinetic energy. The particle takes infinitesimal time $dt$ to dissipate $\frac{d^2s}{dt^2} * ds$ amount of its kinetic energy (assuming the force filed is conservative). So, the rate at which kinetic energy is dissipated by the particle is $\frac{d^2s}{dt^2} * \frac{ds}{dt}$. We will simply write it as $s''s'$, which is the power of the particle according to the notion of Newtonian mechanics.

Now, when the trajectory of the particle becomes the time series (or, rather the graph of the time series) what happens to this dissipated kinetic energy of the particle? Notice, the force is bringing the particle up and down to give 'shape' to its complicated motion plotted as its trajectory, which is very different from the stationary position. Similarly, the kinetic energy is also being dissipated to give shape to the trajectory. In other words, the kinetic energy of the particle is transformed into the 'shape information' or the semantic information (Definition 5) of the trajectory or (the graph of) the time series. Since kinetic energy of the particle is dissipated at the rate of $s''s'$, the information in the form of the shape of the time series is also being encoded at the rate $s''s'$. But $s''s'$ is the kinetic power of the particle at a particular point, therefore its absolute value has been taken as the information power at a particular point. Cumulative information power over a time interval $(a,b)$ is the integration $P_I(s(t)) = \int_a^b |P(s(t))|dt$, as in Definition 6. In Fig 10 spectral power and information power have been compared on epileptic seizure data. Generally both go up during the seizure, but again we have observed information power is having a more uniform trend than the spectral power and therefore like semantic entropy, information power too can be a better biomarker for epileptic seizures. The comparative study between the two measures of power was conducted on 261 focal channels implanted in 21 patients for recording 87 seizures in the Freiburg data set. The outcome has been presented in the histogram plot of Fig 10. In 70% of 87 seizures spectral power and information power show the same trend (in 82% out of the 70% of the cases power goes up and in remaining 18% of the cases it does not go up). In 21% of the seizures information power goes up, whereas spectral power goes down. In remaining 9% of the seizures spectral power goes up, whereas information power goes down.

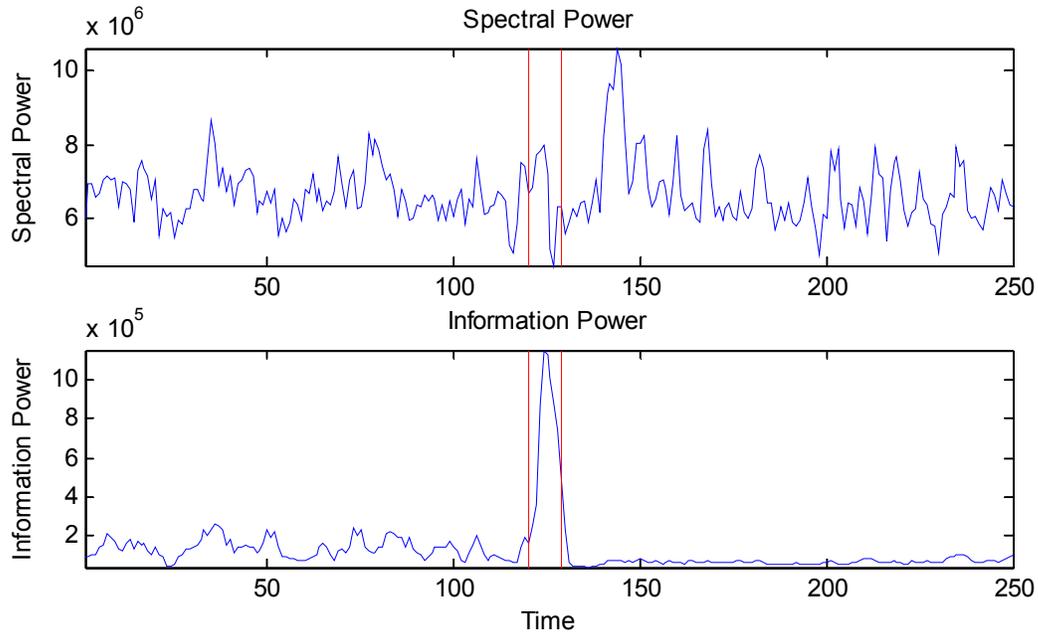

Fig 10. Time versus spectral power (top) and information power (bottom) plot of a human intracranial EEG signal from seizure focus. Vertical lines indicate start and end of an epileptic seizure.

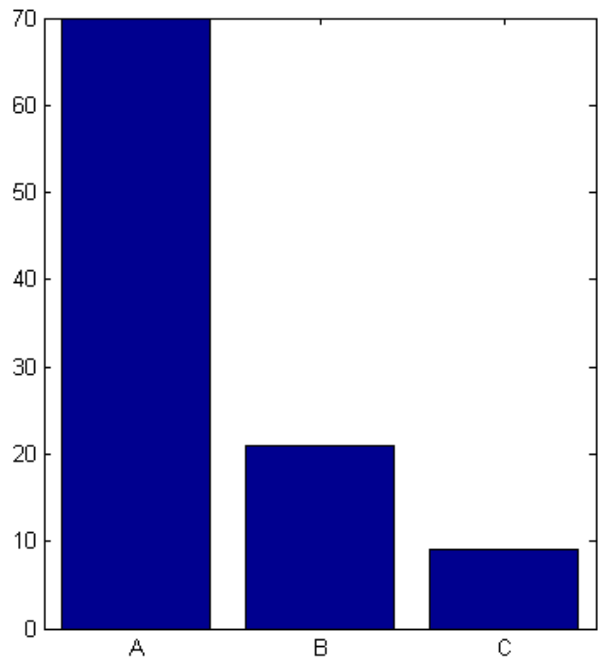

Fig 11. Summary of comparative study between spectral power and information power on 261 focal channels of 21 patients during recording of 87 seizures in Freiburg data set. Histogram A stands for 70% of the cases where both the measures have shown same trend. Histogram B stands for 21% of the cases where information power during the seizure goes up, but spectral power goes down. Histogram C signifies remaining 9% of the cases where during the seizure spectral power goes up but information power goes down.

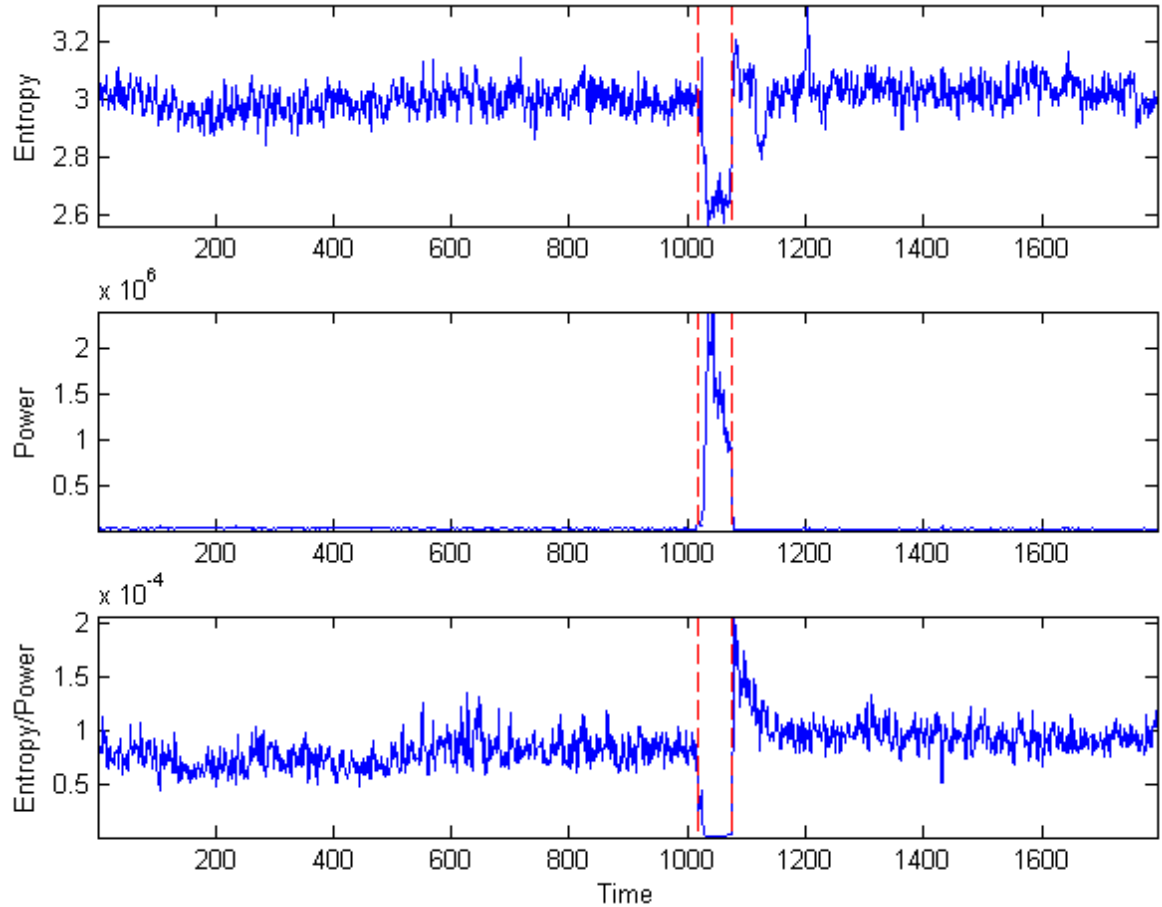

Fig 12. Time versus semantic entropy (top), information power (middle) and semantic entropy / information power (bottom) plots of an intracranial EEG time series from a patient with epilepsy before, during and after a seizure. Seizure start and end are marked with vertical lines. The plots are for one hour duration. Each time point is of 2 seconds duration. During seizure entropy / power almost touches zero for a prolonged duration is the predominant trend observed across the data set (see Table 2).

Since the power goes up and entropy goes down during the seizure, entropy / power being very low can be a potential biomarker for epileptic seizures (Fig 12). We have tested this idea on the Freiburg data set, which is particularly meant for benchmarking new algorithms. The result has been summarized in Table 2. In 72 out of 87 seizures E/P becomes the lowest during seizure as shown in Fig 12. Epileptic seizures have been defined to be highly synchronous actions of the neurons in the brain (Fisher et al. 2005), which is a regular state and therefore entropy is expected to remain low. Since most of the seizures start with low amplitude high frequency oscillations and then gradually progress into high amplitude low frequency waveforms, it is natural that $|P(s(t))|$ will remain high during the seizure. This explains the findings summarized in Table 2. Out of 21 patients in the data base 12 patients showed the trend in 100% cases, that is, for all 53 seizures recorded in 12 patients E/P became lowest during the seizure (in each case seizure has been recorded with 1 hour background signal). For 6 patients the trend is seen in 68% cases, that is 13 out of 19 seizures recorded in those 6 patients showed lowest E/P during seizure (for remaining 6 seizures E/P became lowest somewhere in the background signal). For the remaining 3 patients this trend was seen only in 40% of the recorded seizures.

Table 2
Summary of results of entropy (E) / power (P)
87 seizures have been recorded in 21 patients

| Number of patients | Number of seizures | Percentage of seizures with the trend* |
|---|---|---|
| 12 | 53 | 100% |
| 6 | 19 (One seizure in each patient does not show the trend) | 68% (13 out of 19) |
| 3 | 15 (More than one seizures in each patient does not show the trend) | 40% (6 out of 15) |
| 21 | 87 | |

*Trend means E/P becoming the lowest during seizure

### 3.3 Synchronization

In the previous subsection we have observed that epileptic seizure is a hyper-synchronous phenomenon and that during epileptic seizures E/P goes down significantly in EEG time series compared to the baseline. Intracranial EEG time series is supposed to be superposition of various other electrical time series like synaptic potential, membrane potential, neuronal action potential etc. During seizure these contributing potentials in intracranial EEG become synchronized. The resultant EEG time series becomes regular (therefore the entropy goes down) and intensified (therefore the power goes up). This leads to very low values for E/P when calculated in a windowed manner. Is there any relationship in general between the E/P measure and synchronization of multiple time series then?

When two or more time series are to synchronize they must behave in a similar manner in some sense. In other words, there should be a regular pattern across all the time series of interest. Entropy should go down. The semantic entropy that we have defined here has turned out to be a suitable measure of this phenomenon. On the other hand, information power will go up when the time series will vigorously jitter up and down. Now, if a time series jitters up and down in a random manner there will be no regularity and the entropy will be high. In that case although the power will be high E/P will not be as low. But if a time series jitters up and down vigorously yet in a regular manner its entropy will be low and power will be high leading to a very low E/P. Additionally if that time series is tightly couples to other time series then the former one will try to 'influence' the latter so that the latter time series tend to behave in a similar manner with the former. In other words, the former time series will try to synchronize the latter time series in order to make them evolve the way the former is evolving. So, the coupling strength plays a crucial role in synchronizing a system of several time series. This has been nicely demonstrated in Kuramoto model (Acebron et al. 2005) in which coupling strength is an essential parameter. Kuramoto model is a model of coupled oscillators. Each of the oscillators has a natural or eigen frequency. Since each oscillator is coupled with all other oscillators, every oscillator influences all others and also gets influenced by them. Thus its oscillation in its natural frequency is perturbed and it oscillates in a frequency other than its natural frequency. The entire system of coupled oscillators is modelled by the following equations (one for each $i$).

$$\frac{d\theta_i}{dt} = \omega_i + \frac{K}{N}\sum_{\substack{j=1 \\ j \neq i}}^{N} \sin(\theta_i - \theta_j), \tag{4}$$

where $\theta_i$ is the phase and $\omega_i$ is the natural frequency of the $i$th oscillator respectively, $K$ is the total coupling constant of the system and $N$ is the number of oscillators. Equations (4) represent the basic Kuramoto model, which can be modified and extended in many different ways. One obvious way to extend it to assign the coupling $K_{ij}$ to denote the coupling strength between the $i$th and the $j$th oscillators. Then the total coupling strength of the $i$th oscillator is $K_i = \sum_{\substack{j=1 \\ j \neq i}}^{N} K_{ij}$.

Borrowing terminology from probability theory we can call $K_i$ the *marginal coupling* strength of the $i$th oscillator. We can define a new quantity *synchronizability* $S_i(s_i(t))$ of the $i$th time series $s_i(t)$ in a system of coupled time series as

$$S_i(s_i(t)) = \frac{E_i(s_i(t))}{P_i(s_i(t)) * K_i}, \tag{5}$$

where $E_i$ is the semantic entropy (over an interval) of the $i$th time series and $P_i$ is the information power (over the same interval) and $K_i$ is the marginal coupling strength of the time series. We hypothesize that the time series with the lowest synchronizability during a time interval will influence all other time series in the system to get synchronized with itself during that time interval.

**Conclusion**

Take $n$ consecutive unequal values in a discrete time series and then permute them to get $n!$ different configurations, called ordinal patterns. How these ordinal patterns are occurring in the time series to give it a shape has already been studied well for $n = 3$ (Olofsen et al. 2008, Li and Quyang 2010). Obviously, only 6 ordinal patterns have been considered. Here modelling time series by Newtonian mechanics we have introduced power operator or P-operator and as a result $n = 3$ has become the only valid length of the ordinal patterns. Moreover analysis by P-operator enabled us to identify 13 different shapes for a pattern of length 3 instead of only 6. We could extend this shape analysis to continuous time series as well and identified 32 shapes. A future direction of research will be to investigate if this list of 32 shapes is exhaustive, and if not, what are the other shapes?

Modelling a time series as the trajectory of a particle moving in a force field we have been able to directly relate the 13 configurations of a discrete time series to semantic information encoding into the time series. Semantic information is the information pertaining to the meaning of the time series (on the other hand Shannon information is not at all concerned about the meaning of information, it only pertains to information content) like identifying brain abnormality from EEG signal or heart abnormality from ECG signal. The next challenge will be to relate the 13 symbol strings to the underlying dynamical process responsible for generating the time series.

A new measure of entropy, called semantic entropy (E), has been defined based on the frequency distribution of the 13 configurations for a discrete time series in order to measure regularity of the series. Numerical value of P-operator run on a time series gives the rate at which semantic information is being encoded into the time series. So we call mean absolute value of P-operated time series over an interval as information power (P). When a time series is realized as a superposition of multiple other time series, E goes down as the component time series become more and more synchronous and P goes up. So E/P goes down. This we have observed in real data. Now, consider a network of coupled dynamical systems, each generating a time series. How this network synchronizes is a very important issue in many areas of research (Arenas et al. 2008). Theoretical considerations indicate E/(P*K), where K is coupling constant, should be a measure of synchronization of the whole or part of the network in the sense that more a group of generators are synchronized the more E/(P*K) goes down. An investigation into this aspect can offer us a deeper insight into the synchronization phenomenon in a network of coupled dynamical systems and also can find efficient algorithms to measure synchronizabiltiy.

**Acknowledgement**

This work was supported by a Department of Biotechnology, Government of India grant no. BT/PR7666/MED/30/936/2013. Prof. Subir Kumar Bhandari and Prof. Yogesh Dhandapani are being acknowledged for helpful discussions.